\documentclass[titlepage,aps,pre,twocolumn,amsmath,amssymb,superscriptaddress,reprint,longbibliography,nofootinbib]{revtex4-1}

\usepackage[utf8]{inputenc}
\usepackage{graphicx}
\usepackage{color}
\usepackage[colorlinks=false,citecolor=black]{hyperref}
\usepackage{units}
\usepackage{amsmath}
\usepackage{tabularx}
\usepackage{lipsum}
\usepackage{float}

\usepackage{amssymb}
\usepackage{xcolor}
\usepackage{extarrows}
\usepackage{tikz-cd}
\usepackage{calrsfs}
\usepackage[bottom]{footmisc}

\newcommand{\figref}[1]{Fig.~\ref{#1}}

\DeclareMathOperator{\sign}{sign}

\newcommand{\Ewm}{E_{\text{WM}}}
\newcommand{\Ssh}{S_{\text{Sh}}}
\newcommand{\betabath}{\beta_{\text{bath}}}
\newcommand{\SWM}{S_{\text{WM}}}
\newcommand{\EWM}{E_{\text{WM}}}
\newcommand{\Emax}{E_{\text{max}}}
\newcommand{\bWM}{\beta_{\text{WM}}}
\newcommand{\xbath}{x_{\text{bath}}}
\newcommand{\xWM}{x_{\text{WM}}}
\newcommand{\ignore}[1]{}

\DeclareMathOperator{\artanh}{artanh}

\AtBeginDocument{%
 \newwrite\bibnotes
 \def\bibnotesext{Notes.bib}
 \immediate\openout\bibnotes=\jobname\bibnotesext
 \immediate\write\bibnotes{@CONTROL{REVTEX41Control}}
 \immediate\write\bibnotes{@CONTROL{%
 apsrev41Control,author="08",editor="1",pages="1",title="0",year="1"}}
 \if@filesw
 \immediate\write\@auxout{\string\citation{apsrev41Control}}%
 \fi
}%

\begin{document}

\title{Work extraction from a single negative-temperature bath:
a temperature gradient-driven heat flow}
\author{Wolfgang Rudolf Bauer}
\email{bauer\_w@ukw.de}
\affiliation{
University/University Hospital of W\"urzburg (ZIM),
Oberd\"urrbacher Stra{\ss}e 6,
D-97080 W\"urzburg, Germany, Tel. +49-931-201-39011
}
\altaffiliation[Also at: ]{Comprehensive Heart Failure Center, Am Schwarzenberg 15, A15, D-97080 W\"urzburg, Germany} 

\date{\today}

\begin{abstract}
Work can dissipate as heat within a thermal bath. The reverse process---extracting work from a thermal bath without changing the state of the working system (for instance, without increasing its entropy)---contradicts the second law of thermodynamics. However, if the bath is at negative absolute temperature, this restriction no longer holds in its standard form, which has motivated a modified Kelvin--Planck formulation of the second law. So far, this has sparked a vivid debate, even questioning whether negative absolute temperatures exist in stable systems.
Our approach reconciles the second law of thermodynamics with sole work extraction from a negative-temperature bath. Instead of a single working machine, we consider a large assembly of working systems that can exchange energy via weak thermal coupling. This virtual assembly is assigned a canonical temperature obtained from its Shannon (information) entropy. While this concept may appear unconventional for macroscopic machines, it is natural for molecular-scale working systems. We investigate two paradigmatic examples: a bounded energy ladder with equidistant levels and a freely jointed polymer chain lifting a load. Whenever the canonical temperature of the assembly is cooler than that of the bath, work extraction is simply heat flow from a hot system (a bath at negative temperature) to a cooler system (the assembly of working machines), in full agreement with the second law. This picture also accounts for the counterintuitive ``donkey behavior,'' in which increasing the load prompts a stronger upward response. Finally, we propose a cyclic working machine operating in a single negative-temperature bath.

\begin{description}
\item[PACS numbers] 5.70.Ln, 5.20.Gg
\end{description}            
\end{abstract}
\maketitle

\section{Introduction}
In 1842, Julius Mayer formulated his equivalence principle that work can dissipate into heat \cite{mayer1842}, suggesting that heat is a form of energy that can originate from work. This principle became the foundation of the first law of thermodynamics, i.e., energy conservation, while also pointing to a preferred direction of energy conversion. The reverse process---converting heat into work as a usable form of energy---is conceptually more subtle and motivated the second law of thermodynamics, which introduces entropy and links work extraction to an entropy increase. Among its various formulations, we take the Kelvin--Planck statement as our starting point \cite{thomson1853xv,planck2011vorlesungen}: it is impossible to extract work from a system operating in a cycle solely by cooling a single heat reservoir. Its connection to entropy follows because cyclic operation means that every macroscopic state variable of the working machine (WM), and in particular its entropy, returns to its initial value after one cycle. Extracting work $W>0$\footnote{Work is taken as positive when produced by the working machine. Heat changes $Q$ refer to the bath's energy: heat leaving the bath has a negative sign ($Q<0$), and heat entering it has a positive sign.} from a single reservoir then implies $W+Q=0$ by the first law. For a positive bath temperature $T_{\text{bath}}>0$, heat loss from the bath lowers its entropy by $\Delta S_{\text{bath}}=Q/T_{\text{bath}}<0$. Since the entropy of the WM does not change, the total entropy would decrease, which is forbidden. Thus, the Kelvin--Planck statement implies that a single heat bath at positive temperature cannot yield net work without changing the entropy of the WM\footnote{A heat bath can of course perform work at the expense of increasing a WM's entropy, for example when a gas-filled piston in thermal contact with a bath expands, increases its volume entropy, and lifts a weight.}.

Motivated by the observation of negative spin temperatures \cite{purcell1951nuclear}, Ramsey reformulated the second law for baths at negative temperature \cite{19ramsey56thermodynamics}\footnote{For both positive and negative temperatures, ``hotter'' and ``cooler'' are consistently ordered by $1/T_{\text{hot}}<1/T_{\text{cool}}$.}. In particular, a bath at negative temperature can transfer heat to a WM and thereby enable work production while increasing total entropy; conversely, work cannot simply be dissipated into such a bath without a change of state (entropy) of the WM.

\begin{figure}
\includegraphics[width=9.5cm] {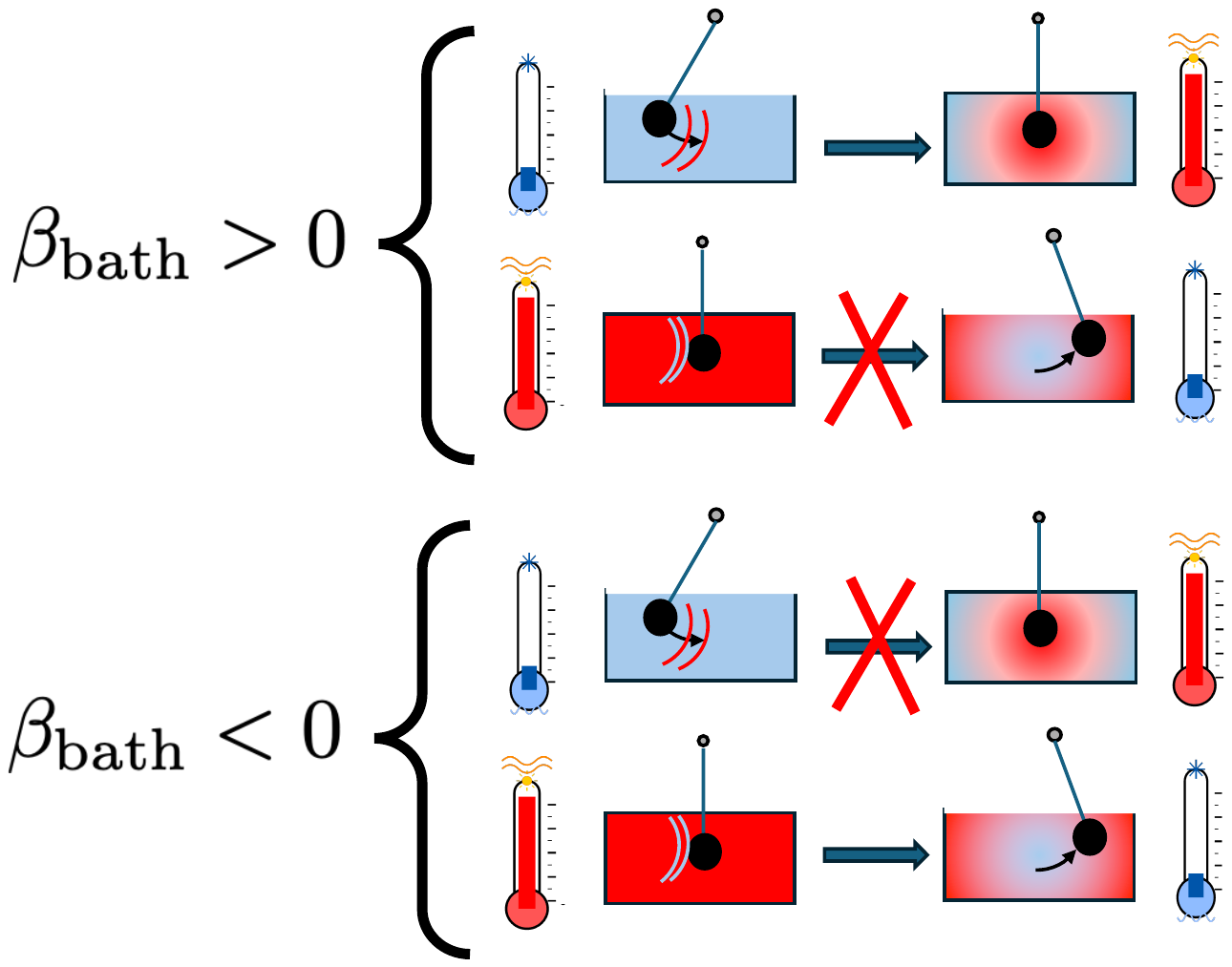}
\caption{Schematic of the relation between working machine, symbolized by a pendulum, and temperature of a bath. For positive temperatures ($\beta_{\text{bath}}=1/T_{\text{bath}}>0 $) work may dissipate to heat, i.e. the working machine  heats up the bath in its vicinity. The reverse process, i.e. pure extraction of work from the bath is impossible. Vice versa a negative temperature bath ($\beta_{\text{bath}}=1/T_{\text{bath}}<0 $) allows the latter, but the 2nd law forbids that the dissipation of work into heat.} 
\label{Fig:1}
\end{figure}

In summary, for a WM in contact with a bath, the two temperature regimes behave oppositely: at positive bath temperatures, work can dissipate into heat in the bath, whereas pure work extraction from the bath is impossible. At negative bath temperatures, work can be extracted from the bath, whereas work cannot simply be converted into heat in the bath (\figref{Fig:1}).

This reformulation of the 2nd law  sounds awkward because it distinguishes between the sign of the  bath's temperature. Instead one would prefer a consistent description of this scenarios in a concise form related to basic thermodynamic quantities. Struchtrup avoided this by denying the ``real'' existence of negative temperatures \cite{struchtrup2018work,struchtrup2019efficiencies}, and instead he considered them as apparent phenomena and as some kind of non-equilibrium parameter. In contrast, our concept accepts negative temperatures of the bath in the same way as positive ones. The main idea is to consider the interaction of bath and working machine, i.e. work dissipation and work extraction in accordance  with Clausius's formulation of the 2nd law of thermodynamics  \cite{clausius1898ueber,clausius1879mechanical}, that direction of heat flow is always from hot to cold. This requires to reformulate dissipation and extraction of work as some heat flow process, i.e. we have to assign a working machine a ``temperature''. In this concept negative temperatures  are consistently included within the framework of thermodynamics, and it reconciles Ramsey's attempt to describe work extraction from negative temperature baths with the 2nd law of thermodynamics.

\section{Weakly coupled working machines as a canonical ensemble}
\label{Sect:WMcanonical}
In conventional thermodynamics, a heat bath consists of a vast number of particles with many uncontrollable microscopic degrees of freedom, whereas a working machine (WM) is a macroscopic device governed by only a few controllable degrees of freedom. For molecular-scale machines, however, this sharp distinction blurs: their characteristic energies and dynamical timescales approach those of the surrounding bath particles, making it natural to assign a temperature to the machine itself.

\begin{figure}
\includegraphics[width=9.0cm]{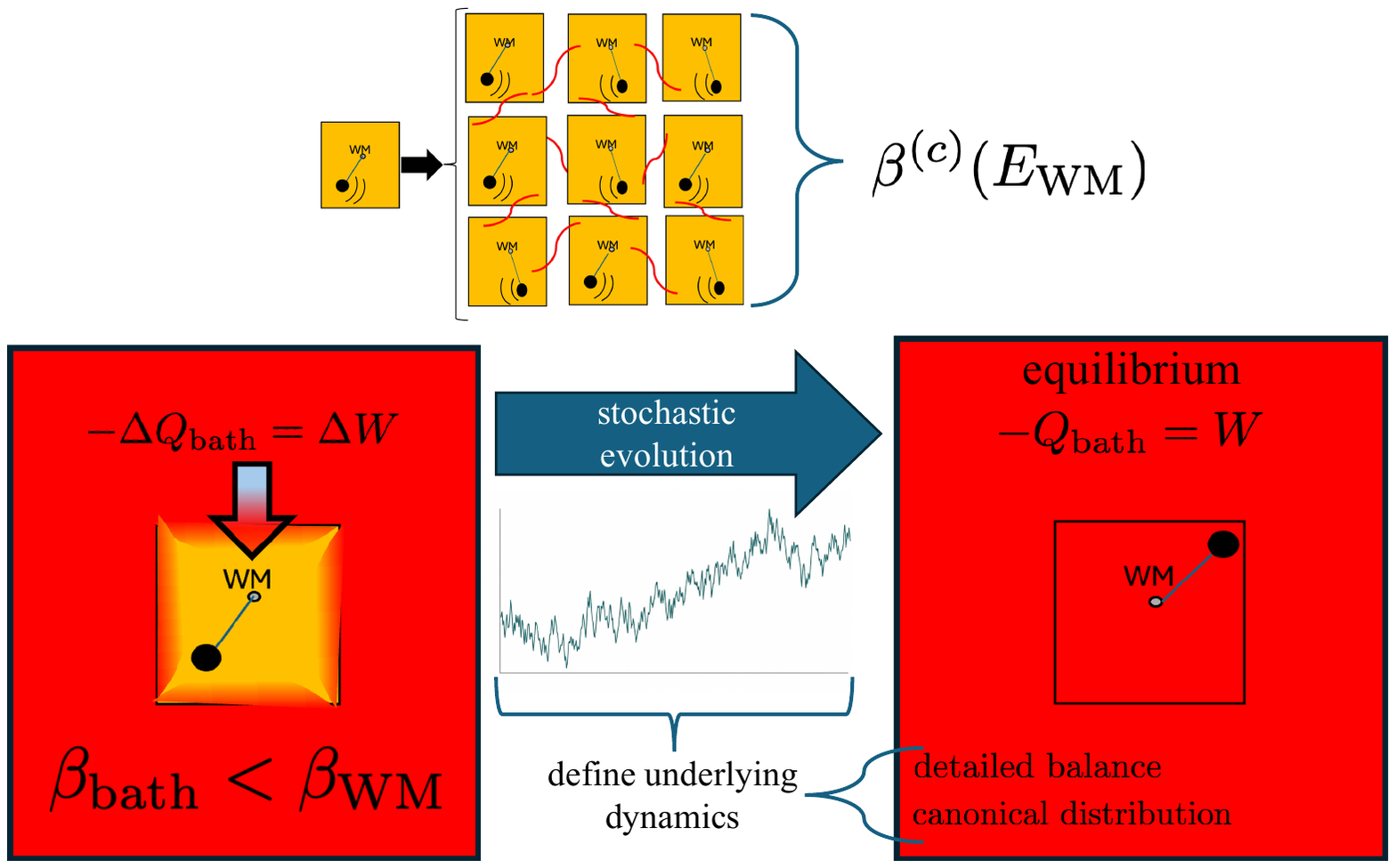}
\caption{Top: Assignment of a canonical temperature to a working machine (WM). A single WM (here a 1D harmonic oscillator) with energy $\Ewm$ is replicated $N$ times, and weak thermal coupling (red wavy lines) allows energy exchange among the copies. The most probable energy distribution defines a canonical inverse temperature $\bWM$ assigned to the WM. Bottom: Work extraction from a hot bath ($\betabath < \bWM$) interpreted as heat transfer down a temperature gradient. Heat $-\Delta Q_{\text{bath}}$ absorbed from the bath is converted into work $\Delta W$, governed by detailed balance and canonical equilibrium fluctuations in the bath.}
\label{Fig:2}
\end{figure}

We first consider working machines with a single degree of freedom, such as a weight-lifting mechanism or a one-dimensional harmonic oscillator. Because such a machine possesses no internal degeneracy, its intrinsic entropy vanishes, $\SWM = 0$, which simplifies the conceptual setup. Consider a single WM with energy $\Ewm$. We mentally replicate this machine $N$ times ($N \gg 1$) and allow weak energetic coupling among the copies so that they can exchange energy (\figref{Fig:2}). By construction, the initial mean energy of this virtual assembly satisfies
\begin{equation}
\bar{E}(0) = \Ewm \;. \label{Eq:energyconstraint}
\end{equation}
Any normalized energy distribution $p(E)$ across the assembly must conserve this mean energy,
\begin{equation}
\sum_{E} p(E) E = \bar{E}(0) \;. \label{Eq:energyconservation}
\end{equation}
For large $N$, the probability $\varpi[p]$ of observing a given energy distribution $p(E)$ scales with its Shannon entropy $\Ssh[p]$ as
\begin{equation}
\varpi[p] \sim \exp\!\big(N\,\Ssh[p]\big) \;, \label{Eq:Shannon}
\end{equation}
where
\begin{equation}
\Ssh[p] = -\sum_{E} p(E)\ln p(E) \;.
\end{equation}
Maximizing $\varpi[p]$---or equivalently $\Ssh[p]$---with respect to $p(E)$ under the energy-conservation constraint~\eqref{Eq:energyconservation}\footnote{The normalization constraint $\sum_{E} p(E) = 1$ merely contributes an overall multiplicative constant that cancels out in Eq.~\eqref{Eq:canonical1}.} via a Lagrange multiplier $\beta$ yields the stationarity condition
\begin{equation}
\nabla_p \Ssh[p_c] - \beta E = 0 \;.
\end{equation}
This directly leads to the canonical distribution
\begin{equation}
p_c(E) = \frac{e^{-\beta E}}{Z(\beta)} \;, \label{Eq:canonical1}
\end{equation}
with the partition function $Z(\beta) = \sum_{E} e^{-\beta E}$. By analogy with standard canonical statistical mechanics, we identify the Lagrange multiplier $\beta$ with the inverse temperature of the WM assembly (in units where $k_{\text{B}} = 1$),
\begin{equation}
\frac{1}{T} = \beta \;.
\end{equation}
Since the mean energy of the canonical distribution is $-\partial_{\beta}\ln Z(\beta)$, Eq.~\eqref{Eq:energyconstraint} uniquely relates the energy $\Ewm$ of the single working machine to its assigned inverse temperature $\bWM$ via
\begin{equation}
\Ewm(\bWM) = -\partial_{\beta}\ln Z(\beta)\big|_{\beta=\bWM} \;. \label{Eq:EWMPartition}
\end{equation}

Within this framework, work dissipation and work extraction in a thermal bath are both governed by the temperature gradient between the bath and the WM. If the bath is cooler than the WM ($\bWM < \betabath$), energy flows from the WM into the bath and dissipates as heat. Conversely, if the bath is hotter than the WM ($\bWM > \betabath$), heat flows from the bath to the WM, where it is converted into work (\figref{Fig:2}).

\section{Work extraction from a bath at negative temperature}
\label{Sect:WorkExtract}

\subsection{The bath with negative temperature}
As a bath capable of exhibiting negative temperatures, we consider a paramagnetic spin-$1/2$ gas in a homogeneous external magnetic field $B$ aligned along the $z$ axis. With gyromagnetic ratio $\gamma$, the energy of a single spin is
\begin{equation}
E_{\text{spin}} = -s_z \hbar\gamma B \;.
\end{equation}
The inverse temperature $\betabath$ of such a paramagnetic spin gas is positive when more spins align parallel to the external field ($\langle E_{\text{spin}}\rangle < 0$) and negative in the opposite case ($\langle E_{\text{spin}}\rangle > 0$), so that
\begin{equation}
\sign(\betabath) = -\sign(\langle E_{\text{spin}}\rangle) \;.
\end{equation}

\subsection{Temperature of a working machine with equidistant energy spacing}
For the working machine, we assume a bounded energy spectrum consisting of $M$ non-degenerate levels with equidistant spacing $\Delta E$,
\begin{equation}
E_m = (m-1)\Delta E \;, \quad 1 \le m \le M \;, \label{Eq:WMequidistant}
\end{equation}
which can be realized, for example, by a bounded harmonic oscillator or a weight lifted in a constant force field. Using Eq.~\eqref{Eq:EWMPartition}, the assigned inverse temperature $\bWM$ of the WM is determined by
\begin{align}
\Ewm &= \tfrac{\Emax}{2} + \nonumber \\
&\quad \tfrac{\Delta E}{2} \left[ \coth\!\left(\tfrac{\bWM \Delta E}{2}\right) - M \coth\!\left(M \tfrac{\bWM \Delta E}{2}\right) \right] \;, \label{Eq:Ewm}
\end{align}
with $\Emax = (M-1)\Delta E$. The sign of $\bWM$ is governed by the energy relative to the midpoint of the spectrum,
\begin{equation}
\bWM \;
\begin{cases}
> 0 & \text{if } \Ewm < \Emax/2 \;, \\
< 0 & \text{if } \Ewm > \Emax/2 \;.
\end{cases}
\end{equation}
A bounded energy spectrum allows the WM itself to reach negative temperatures. In terms of heat transfer from a hot bath to a cooler WM, this enables the WM to equilibrate with a negative-temperature bath and reach the same canonical temperature.

\subsection{Working machine in contact with a bath at negative temperature}
\label{Subsection:WMnegativetemp}
Placing this working machine in a bath at negative temperature with weak energetic coupling induces stochastic transitions across its energy levels. We model these transitions as a stationary Markov process restricted to nearest-neighbor hops.

\begin{figure}
\includegraphics[width=8.5cm]{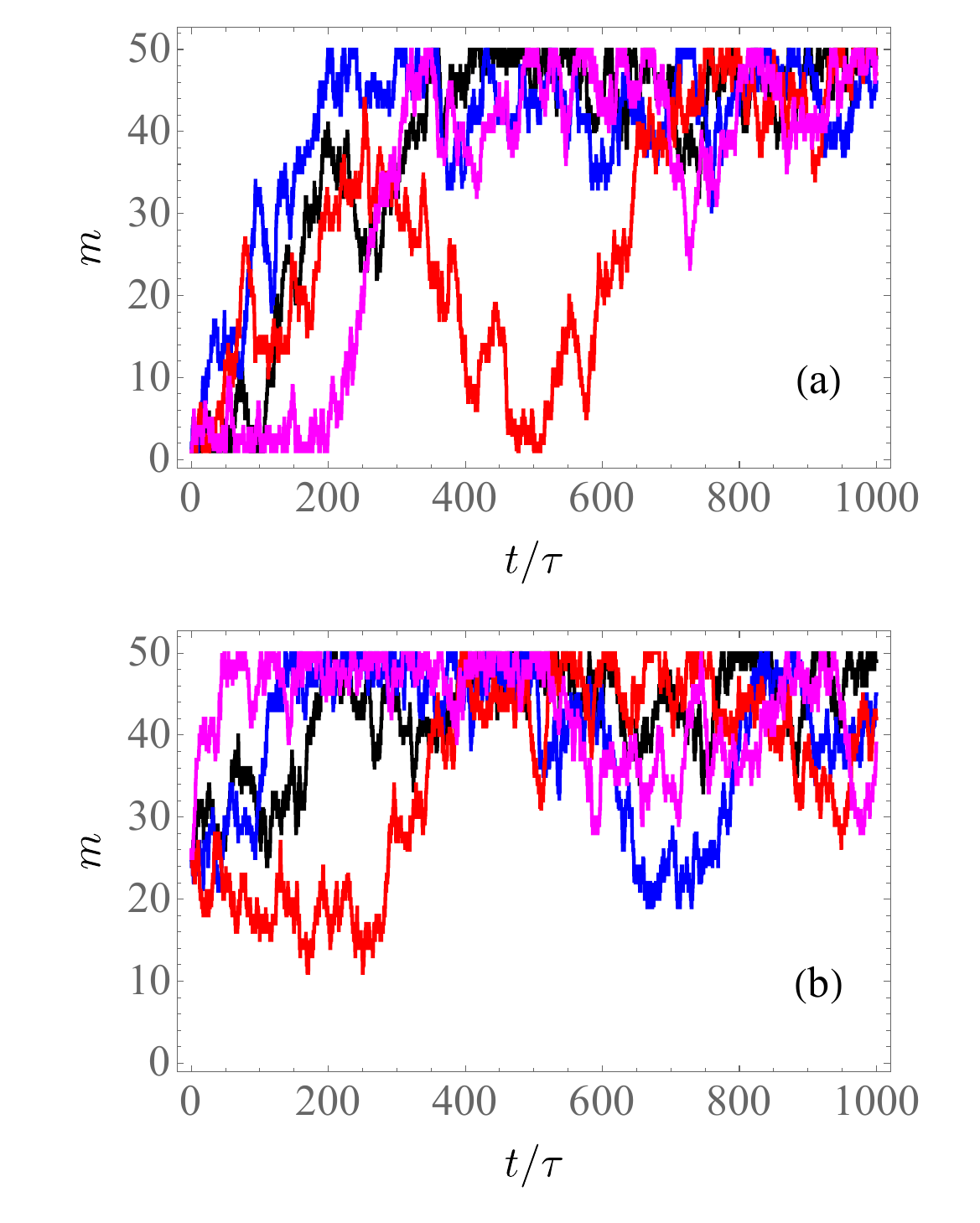}
\caption{Four stochastic trajectories of working machines with $M=50$ equidistant energy levels [Eq.~\eqref{Eq:WMequidistant}] in contact with a negative-temperature bath ($\betabath\Delta E = -0.1$). Trajectories start either from the ground state $m=1$ ($\bWM = +\infty$) or from the middle of the spectrum at $m=25$ ($\bWM\Delta E \approx 2.4 \times 10^{-3}$; see text). Transition rates follow Eq.~\eqref{Eq:transitionrates}.}
\label{Fig:3}
\end{figure}
\begin{figure}
\includegraphics[width=8.5cm]{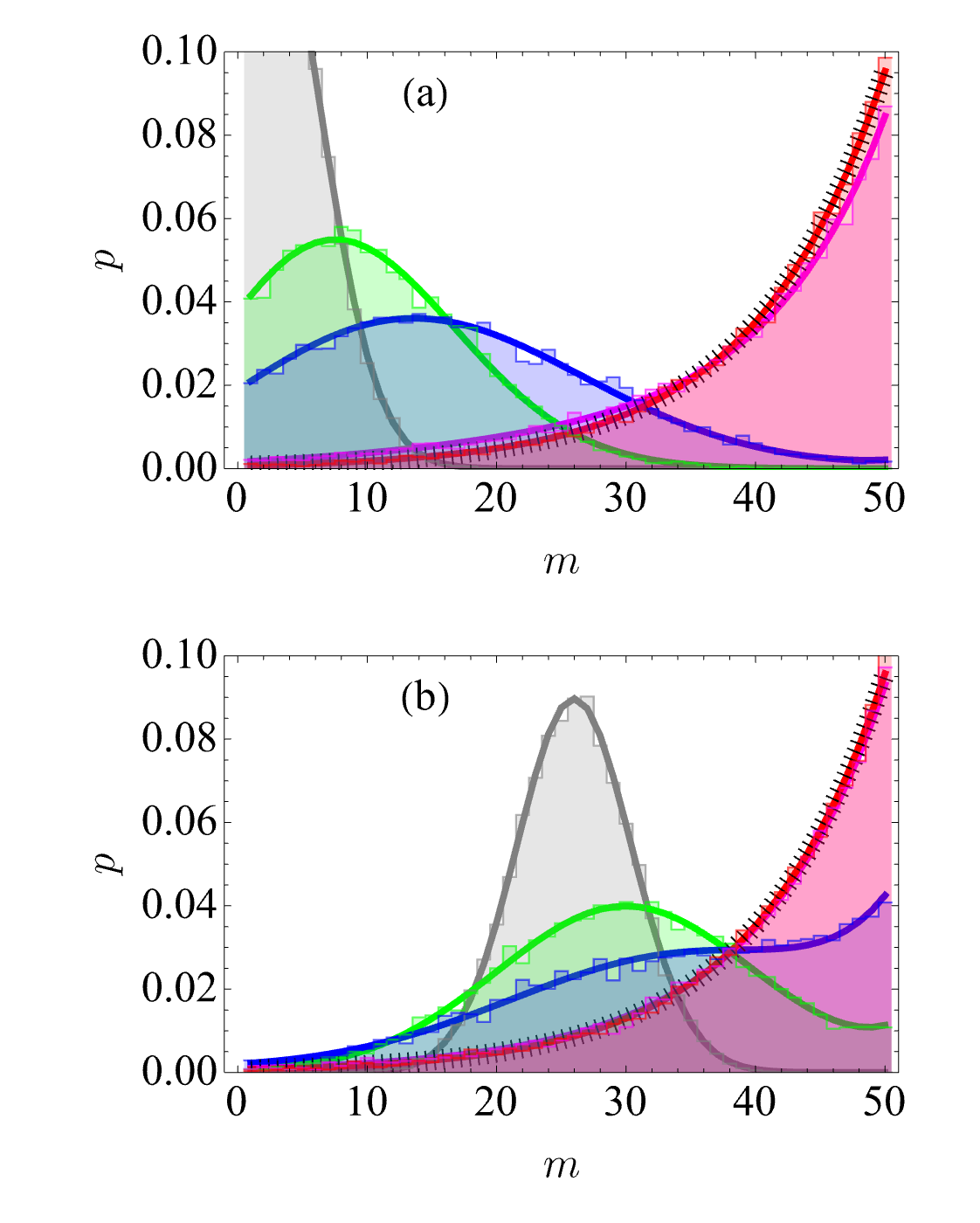}
\caption{Time evolution of the energy distribution of working machines starting at (a) the ground state $m=1$ and (b) $m=25$ at $t/\tau = 10$ (gray), $50$ (green), $100$ (blue), $500$ (magenta), and $1000$ (red). Histograms are obtained from $10\,000$ stochastic trajectories as in \figref{Fig:3}; solid curves show the analytical Green's function $p(m'|m;t) = \big(\exp(\mathbf{R}t)\big)_{m',m}$ (Appendix~\ref{Appendix2}). The red dotted curve indicates the equilibrium distribution $p^{(\text{eq})}(m)$ [Eq.~\eqref{Eq:equilibriumdistribution}] reached as $t \to \infty$.}
\label{Fig:4}
\end{figure}
\begin{figure}
\includegraphics[width=8.5cm]{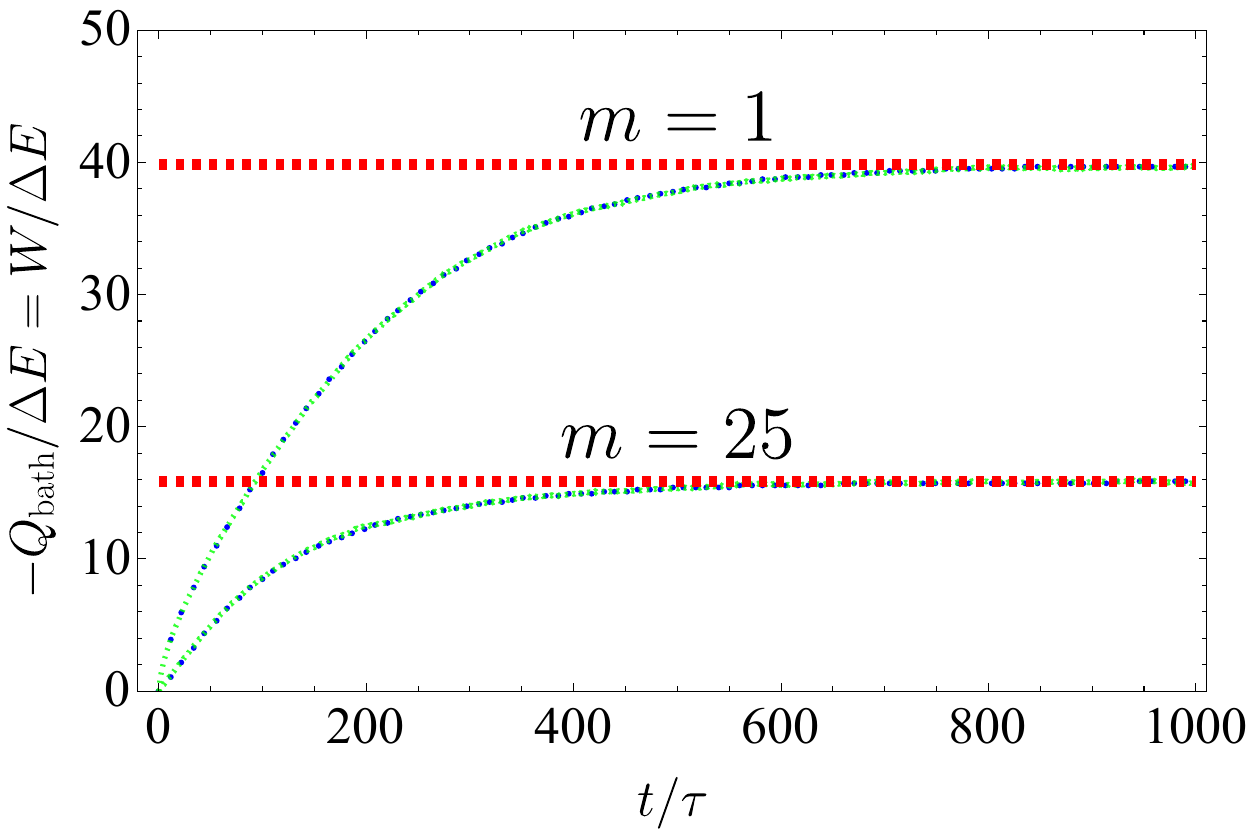}
\caption{Mean heat $-Q_{\text{bath}}$ absorbed from the bath and converted into work $W$ for WMs starting at $m=1$ or $m=25$. Blue circles/dots (as plotted) show results from $10\,000$ stochastic trajectories, overlaid by the analytical Green's function result [Eq.~\eqref{Eq:Green}]. The red dotted line marks the asymptotic work $W_{\infty} = \bar{E}(t\to\infty) - \Ewm$.}
\label{Fig:5}
\end{figure}

To determine the transition rates, we first consider the long-time limit in which the WM assembly equilibrates with the bath at inverse temperature $\betabath$ (\figref{Fig:2}). Because the energy spectrum is finite and bounded from above, equilibration is guaranteed. In equilibrium, detailed balance (microreversibility) holds, meaning that joint probabilities for forward and reverse transitions over a time interval $t$ are equal,
\begin{equation}
p^{(\text{eq})}_{\betabath}(m_2,m_1;t) = p^{(\text{eq})}_{\betabath}(m_1,m_2;t) \;. \label{Eq:detailedbalance1}
\end{equation}
Writing the joint probability as $p^{(\text{eq})}_{\betabath}(m_2,m_1;t) = p_{\betabath}(m_2|m_1;t)\,p^{(\text{eq})}_{\betabath}(m_1)$ and using the canonical equilibrium distribution $p^{(\text{eq})}_{\betabath}(E_m) \propto e^{-\betabath E_m}$ at negative $\betabath$\footnote{For negative temperatures, the canonical form follows, for instance, by expanding the bath entropy with respect to the energy change $\Delta E_{\text{bath}} = -E_m$ \cite{puglisi2017temperature,baldovin2021statistical}, or by maximizing the total Shannon entropy of bath and WM assembly; the sign of the resulting temperature is determined by the mean energy.}, the nearest-neighbor transition rates
\begin{equation}
r_{m\pm 1, m} = \lim_{t\to dt} \frac{p_{\betabath}(m\pm 1|m; t)}{t}
\end{equation}
satisfy
\begin{equation}
\frac{r_{m\pm 1, m}}{r_{m, m\pm 1}} = e^{\mp \betabath \Delta E} \;, \label{Eq:detailedbalance2}
\end{equation}
just as for positive temperatures. We choose the symmetric Glauber-type rates \cite{Nadler86}
\begin{equation}
r_{m\pm 1, m} = \tau^{-1} \exp\!\left(\mp \frac{\betabath \Delta E}{2}\right) \;, \label{Eq:transitionrates}
\end{equation}
with a characteristic rate $\tau^{-1}$. For $\betabath < 0$, upward transitions in energy are favored over downward ones. In short, the transition dynamics of the WM is fixed by microreversibility and canonical fluctuations in the final equilibrium state (\figref{Fig:2}).

We analyze this stochastic dynamics both numerically and analytically, measuring time in units of $\tau$. The transition rates in Eq.~\eqref{Eq:transitionrates} describe a biased random walk (discrete diffusion) on a linear potential $E_m = (m-1)\Delta E$, corresponding to a constant driving force proportional to $-\Delta E$. Numerical simulations follow the standard procedure outlined in Appendix~\ref{Appendix1}. Analytically, the rates form the tridiagonal rate matrix $\mathbf{R}$ (Appendix~\ref{Appendix2}), and the master equation yields the propagator (Green's function)
\begin{equation}
p(m'|m;t) = \big(\exp(\mathbf{R}t)\big)_{m',m} \;.
\end{equation}
Spectral decomposition of $\mathbf{R}$ gives
\begin{equation}
p(m'|m;t) = p^{(\text{eq})}(m') + f(m',m,t) \;, \label{Eq:Green}
\end{equation}
with the canonical equilibrium distribution
\begin{equation}
p^{(\text{eq})}(m') \propto e^{-\betabath E_{m'}} \label{Eq:equilibriumdistribution}
\end{equation}
and a multi-exponential relaxation term $f(m',m,t)$ (Appendix~\ref{Appendix2}).

As a concrete example with $M=50$ levels and bath parameter $\betabath\Delta E = -0.1$, we consider two initial states: first, the ground state $m=1$ ($\Ewm = 0$), which corresponds to $\bWM = +\infty$ ($T_{\text{WM}} = +0$) and thus a very ``cool'' initial state; second, the middle level $m=25$ ($\Ewm = 24\Delta E$), where Eq.~\eqref{Eq:Ewm} in linear approximation gives $\bWM\Delta E = -6/(1-M^2) \approx 2.4 \times 10^{-3}$, i.e., a much higher initial energy yet still $\bWM > \betabath$ (cooler than the bath).

The sample trajectories in \figref{Fig:3} illustrate the drift toward higher energies as heat is drawn from the bath and converted into work. \figref{Fig:4} shows how the distribution $p(m'|m,t)$ evolves from $m=1$ and $m=25$ toward the common equilibrium distribution $p^{(\text{eq})}(m') = \lim_{t\to\infty} p(m'|m,t)$ [Eq.~\eqref{Eq:equilibriumdistribution}]. The absorbed heat converted into work,
\begin{equation}
W(t) = -Q(t) = \sum_{m'} (m'-1)\Delta E \, p(m'|m,t) - \Ewm \;,
\end{equation}
is plotted in \figref{Fig:5}. Because $\betabath < \bWM$ in both cases, work extraction corresponds to heat flow from the hotter negative-temperature bath to the cooler WM assembly. Owing to the upper energy bound, the assembly eventually equilibrates with the bath as $t\to\infty$, approaching the asymptotic work $W_{\infty} = \bar{E}(\infty) - \Ewm$ (red dotted line in \figref{Fig:5}), where $\bar{E}(\infty)$ follows from Eq.~\eqref{Eq:Ewm} with $\bWM$ replaced by $\betabath$.

\section{Working machine with internal degrees of freedom}

\subsection{Definition of temperature of the working machine}
The previous example of a working machine with a bounded, equidistant energy spectrum had no internal degrees of freedom, i.e., its energy levels were non-degenerate. We now lift this restriction and allow multiple microstates to share the same energy. This degeneracy gives rise to a Boltzmann entropy of a WM at energy $E$,
\begin{align}
\SWM(E) &= \ln \mathfrak{n}(E) \;, \label{Eq_app:entropyWM} \\
\mathfrak{n}(E) &= \text{number of microstates with energy } E \;. \nonumber
\end{align}
As in Sects.~\ref{Sect:WMcanonical} and~\ref{Sect:WorkExtract}, we start from a WM with initial energy $\Ewm$, replicate it $N$ times, and introduce weak thermal coupling among the copies. After equilibration, the assembly settles into a canonical energy distribution, whose inverse temperature is assigned to the WM at energy $\Ewm$. Unlike in Sect.~\ref{Sect:WMcanonical}, the most probable distribution $p_c(E)$ is now obtained by maximizing the negative Kullback--Leibler divergence (Appendix~\ref{Appendix3}),
\begin{equation}
S_{\text{KL}}[p] = -\sum_{E} p(E) \ln\!\left(\frac{p(E)}{e^{-\SWM(E)}}\right) \;,
\end{equation}
with respect to $p(E)$ under the mean-energy constraint. This yields the canonical distribution
\begin{equation}
p_c(E) = \frac{e^{\SWM(E) - \beta E}}{Z(\beta)} \;, \label{Eq:canonical2}
\end{equation}
with inverse temperature $\beta$ and partition function
\begin{equation}
Z(\beta) = \sum_{E} e^{\SWM(E) - \beta E} \;. \label{Eq:partitionsum2}
\end{equation}
Because the mean energy $-\partial_{\beta}\ln Z(\beta)$ of the assembly equals $\Ewm$ by construction, the assigned inverse temperature $\bWM$ of the WM satisfies
\begin{equation}
\Ewm(\bWM) \overset{!}{=} -\partial_{\beta}\ln Z(\beta)\big|_{\beta=\bWM} \;. \label{Eq:betaWM2}
\end{equation}

\subsection{The freely jointed chain model}
\begin{figure}
\includegraphics[width=8.5cm]{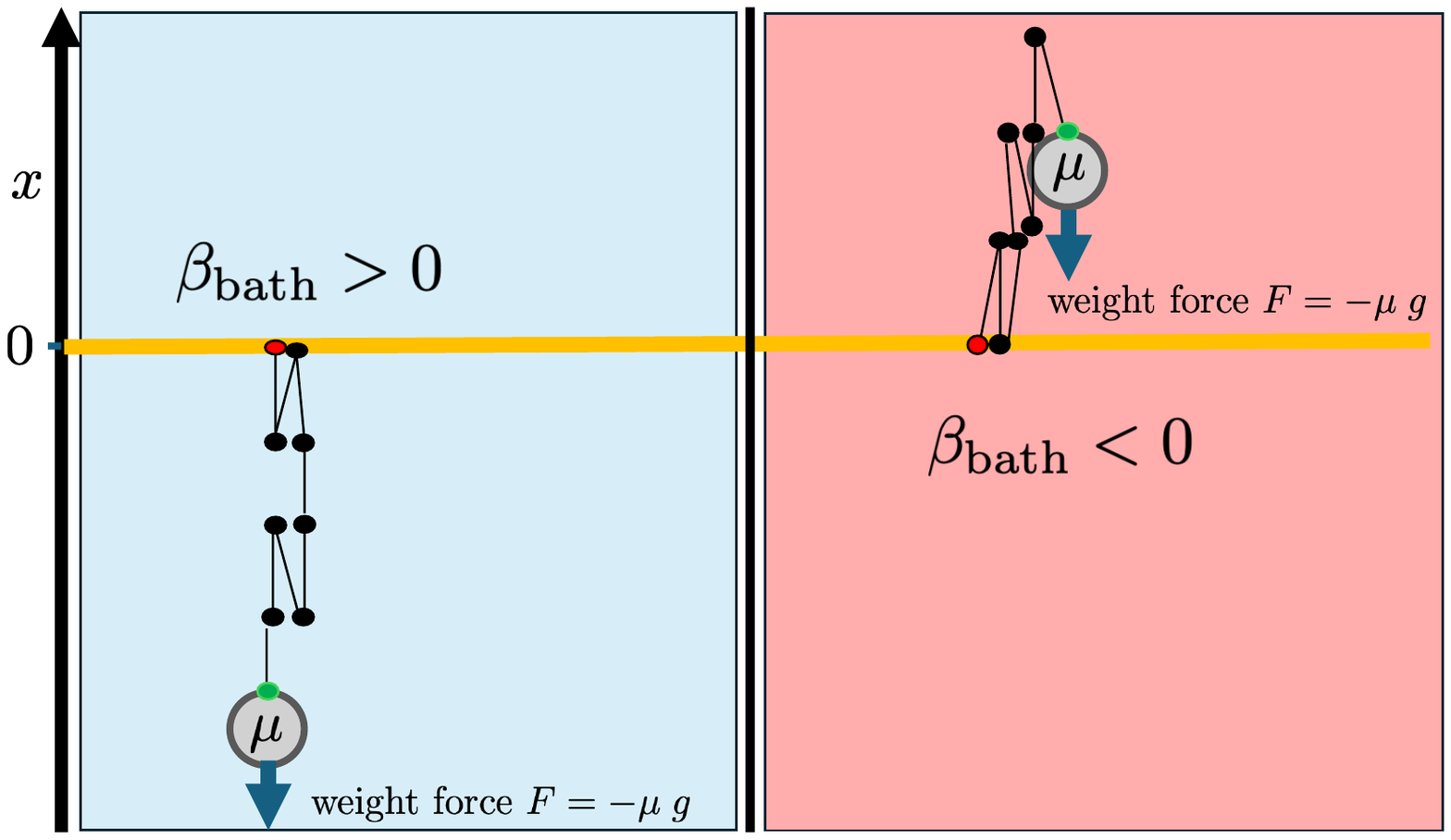}
\caption{Schematic of the freely jointed chain in a bath at positive (left) and negative (right) temperature. One end is fixed at $x=0$, while the free end carries a mass $\mu$ subject to a downward force $F = -\mu g$ ($g>0$). At positive bath temperature the mass settles below $x=0$; at negative bath temperature it is lifted above $x=0$.}
\label{Fig:6}
\end{figure}

A prototypical working machine with internal degrees of freedom is the freely jointed chain, a standard model for entropic elasticity in polymers. In its simplest one-dimensional version along the $x$ axis, the chain consists of $M$ segments of equal length $l$, each oriented either parallel (up, $+x$) or antiparallel (down, $-x$) to the axis. One end is fixed at $x=0$, while the free end experiences a downward force $F = -\mu g$ (with $g>0$), realized in \figref{Fig:6} by a mass $\mu$ in a homogeneous gravitational field. For $u$ segments pointing up and $d = M - u$ pointing down, the chain length is
\begin{equation}
x = (u - d)l = m\,l \;, \label{Eq:length}
\end{equation}
where $m = u - d$. The corresponding potential energy is
\begin{equation}
E = -x F = (u - d)\Delta E \;, \label{Eq:EnergyWM2}
\end{equation}
with the single-segment energy increment
\begin{equation}
\Delta E = -l F = l\mu g \;.
\end{equation}
The number of microstates with fixed $u$ (and hence energy $E$) is $\binom{M}{u}$, giving the WM entropy
\begin{equation}
\SWM(E) = \ln\binom{M}{u} = \ln\binom{M}{\tfrac{M-m}{2}} \;. \label{Eq:EntropyChain}
\end{equation}
Using Eq.~\eqref{Eq:EnergyWM2} and the binomial expansion, the partition function in Eq.~\eqref{Eq:partitionsum2} evaluates to
\begin{align}
Z(\beta) &= \sum_{u=0}^{M} \binom{M}{u} e^{-\beta(2u-M)\Delta E} \nonumber \\
&= e^{\beta \Emax}\left(1 + e^{-2\beta\Delta E}\right)^M \;,
\end{align}
where $\Emax = M\Delta E = -LF$ is the energy at full extension $L = Ml$ along the $x$ axis. The mean energy is
\begin{align}
\bar{E} &= -\partial_{\beta}\ln Z(\beta) \nonumber \\
&= -\Emax \tanh(\beta\Delta E) \;. \label{Eq:meanenergyfreejointchain}
\end{align}
From Eq.~\eqref{Eq:betaWM2}, a WM at initial position $\xWM$ with energy $\EWM = -\xWM F$ is assigned the inverse temperature
\begin{align}
\bWM &= -\frac{1}{\Delta E}\artanh\!\left(\frac{\EWM}{\Emax}\right) \nonumber \\
&= -\frac{1}{\Delta E}\artanh\!\left(\frac{\xWM}{L}\right) \nonumber \\
&\approx -\frac{1}{\Delta E}\left(\frac{\xWM}{L}\right) = \frac{1}{lF}\left(\frac{\xWM}{L}\right) \;, \label{Eq:bWM2II}
\end{align}
where the linear approximation holds for $|\xWM/L| \lesssim 0.5$. Thus, $\bWM > 0$ when the load hangs below the anchor ($\xWM < 0$), and $\bWM < 0$ when it lies above ($\xWM > 0$).

\subsection{The freely jointed chain in contact with a heat bath at negative temperature}

\begin{figure}
\includegraphics[width=8.5cm]{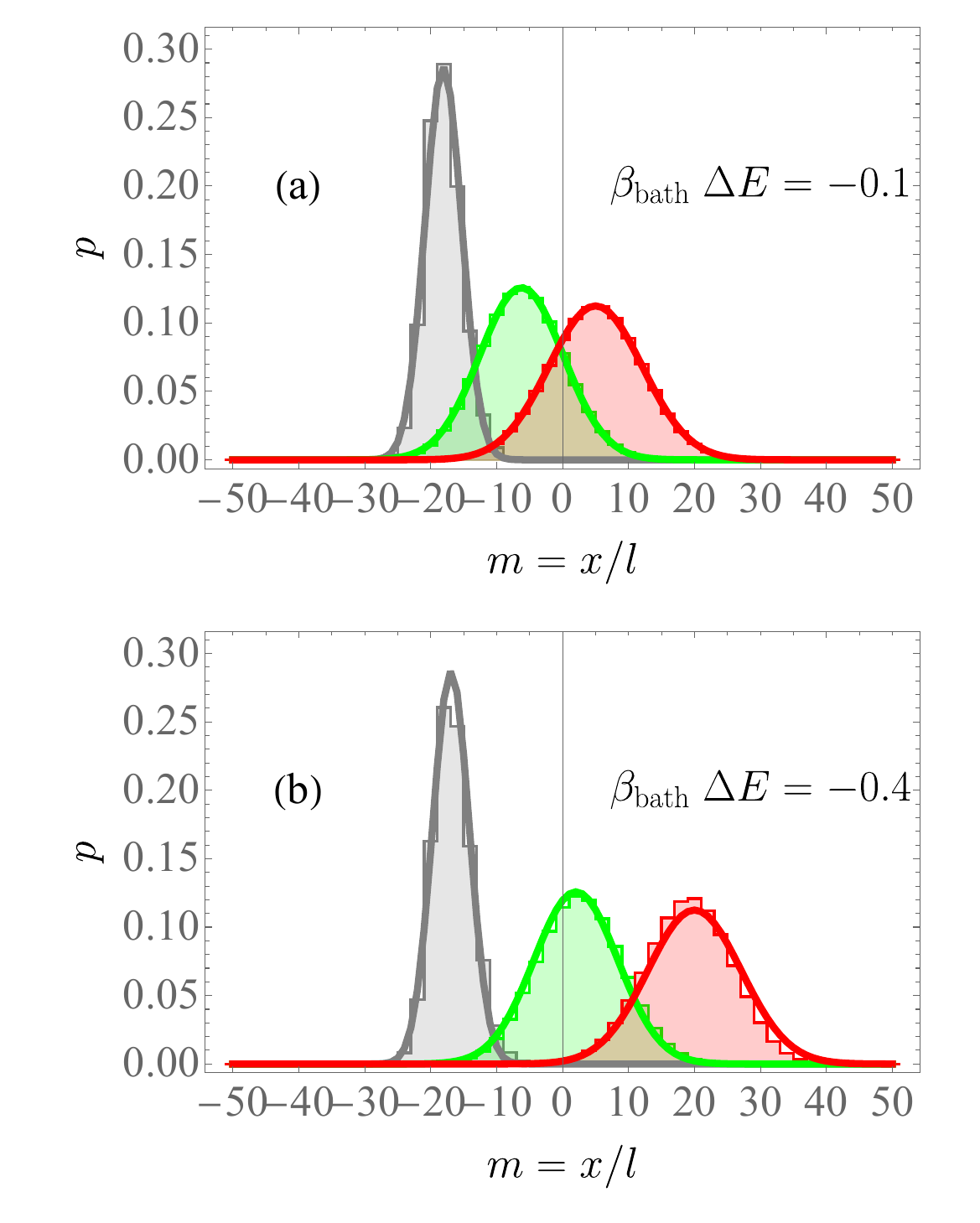}
\caption{Conditional probability $p$ [Eq.~\eqref{Eq:Green2}] of finding the free end of a chain with $M=50$ segments at position $m = x/l$ for times $t/\tau = 1$ (gray), $10$ (green), and $300$ (red), starting from $\xWM = -20l$ ($\bWM\Delta E = \frac{1}{2}\ln(7/3) > 0$) under a load force $F < 0$ and negative bath temperatures $\betabath\Delta E = -0.1$ (a) and $-0.4$ (b). Histograms show the exact discrete Markov Green's function [Eqs.~\eqref{Eq:MasterChain} and~\eqref{Eq:Green2}]; solid curves show the Gaussian approximation [Eq.~\eqref{Eq:GaussGreen3}]. By $t/\tau = 300$, the system nearly reaches equilibrium centered at $\xbath = 5l$ in (a) and $\xbath = 20l$ in (b) [Eq.~\eqref{Eq:xbath}].}
\label{Fig:7}
\end{figure}
\begin{figure}
\includegraphics[width=8.5cm]{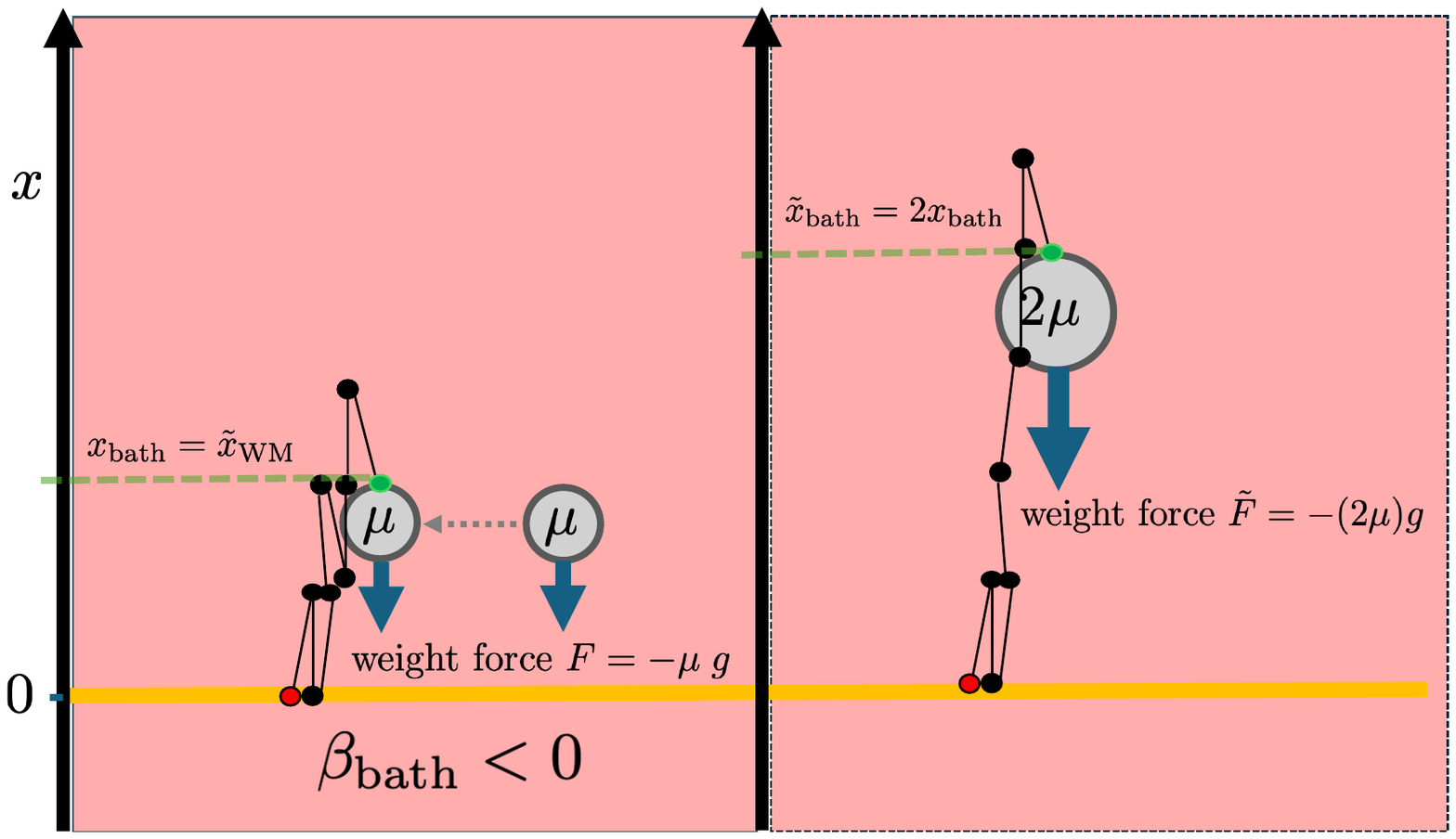}
\caption{Left: A mass $\mu$ attached to the free end is lifted to $\xbath > 0$ in a negative-temperature bath, where a second mass $\mu$ is added. Right: Although the downward load doubles, the combined mass $2\mu$ is lifted further to $\tilde{x}_{\text{bath}} = 2\xbath$.}
\label{Fig:8}
\end{figure}
\begin{figure}
\includegraphics[width=8.5cm]{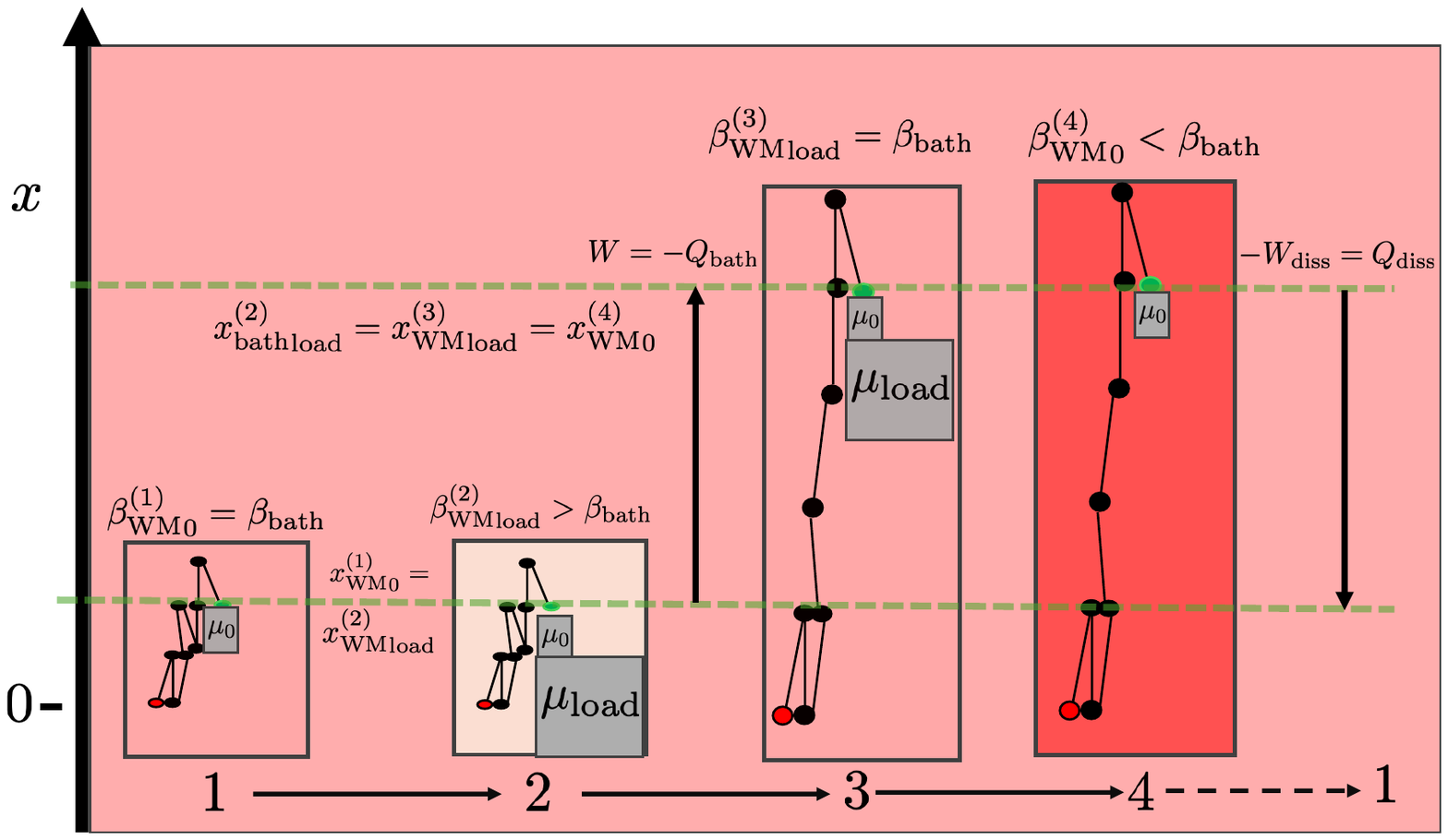}
\caption{Four-step cyclic process for extracting work from a single negative-temperature bath. State~1: WMs with baseline mass $\mu_0$ are at their bath equilibrium position ${\xWM^{(1)}}_0 = {x_{\text{bath}}}_0$. State~2: An additional load mass $\mu_{\text{load}}$ is attached at ${\xWM^{(2)}}_{\text{load}} = {\xWM^{(1)}}_0$. State~3: Absorbing heat $Q_{\text{bath}} < 0$ from the bath lifts the combined mass $\mu_0 + \mu_{\text{load}}$ (doing work $W$) to the new equilibrium position ${x^{(2)}_{\text{bath}}}_{\text{load}}$. State~4: The load mass is removed, making the WMs hotter than the bath (${\bWM^{(4)}}_0 < \betabath$) so they sink back to ${\xWM^{(1)}}_0$ while dissipating heat $Q_{\text{diss}} > 0$ into the bath, returning to state~1.}
\label{Fig:9}
\end{figure}

As in Sect.~\ref{Subsection:WMnegativetemp}, contact with the bath induces Markovian transitions between nearest-neighbor states labeled by $m = x/l$ [Eq.~\eqref{Eq:length}]. Flipping a single segment changes $m$ by $|\Delta m| = 2$ and the energy by $|2\Delta E|$. Detailed balance at inverse bath temperature $\betabath$ requires
\begin{equation}
\frac{r_{m\pm 2, m}}{r_{m, m\pm 2}} = \frac{p^{(\text{eq})}_{\betabath}(m\pm 2)}{p^{(\text{eq})}_{\betabath}(m)} \;. \label{Eq:detailedbalance3}
\end{equation}
Substituting the canonical distribution [Eq.~\eqref{Eq:canonical2}] and the entropy [Eq.~\eqref{Eq:EntropyChain}] gives
\begin{align}
\frac{r_{m\pm 2, m}}{r_{m, m\pm 2}} &= \frac{e^{\SWM(m\pm 2)}}{e^{\SWM(m)}} e^{\mp 2\betabath\Delta E} \nonumber \\
&= \exp\!\left(\underbrace{\Delta S_{m\pm 2, m} + \overbrace{\betabath(\mp 2\Delta E)}^{=\Delta S_{\text{bath}}}}_{=\Delta J = -\betabath\Delta A}\right) \label{Eq:Massieu} \\
&= \frac{M \mp m}{M \pm m + 2}\, e^{\mp 2\betabath\Delta E} \;, \label{Eq:ratequotient}
\end{align}
where $\Delta S_{m\pm 2, m}$ is the WM entropy change and $\Delta J$ (linked to the free-energy change $\Delta A$) is the change in the Massieu potential. Following \cite{Nadler86}, we choose the transition rates
\begin{equation}
r_{m\pm 2, m} = \tau^{-1} \sqrt{\frac{M \mp m}{M \pm m + 2}\, e^{\mp 2\betabath\Delta E}} \;, \label{Eq:MasterChain}
\end{equation}
with characteristic timescale $\tau$. These rates define the rate matrix $\mathbf{R} = (r_{m_2, m_1})$, with diagonal entries $r_{m,m} = -(r_{m+2,m} + r_{m-2,m})$ enforcing probability conservation (modified appropriately at the boundaries $\pm M$). Starting from $m_1$, the conditional probability (Green's function) to reach $m_2$ after time $t$ is
\begin{equation}
\big(p_{\betabath}(m_2|m_1;t)\big) = \exp(\mathbf{R}t) \;. \label{Eq:Green2}
\end{equation}
The drift direction of the free end is governed by $\Delta J$ (or $\Delta A$), which combines the internal entropic pull toward $x = 0$ (where $\SWM$ is maximal) with the bath-mediated contribution $\Delta S_{\text{bath}} = \betabath\Delta Q_{\text{bath}}$. Movement along the downward force $F$ transfers heat into the bath ($\Delta Q_{\text{bath}} > 0$), whereas upward movement extracts heat ($\Delta Q_{\text{bath}} < 0$). For $\betabath > 0$, $\Delta S_{\text{bath}} > 0$ favors downward motion ($m \to m-2$), pulling the weight below $x=0$; for $\betabath < 0$, $\Delta S_{\text{bath}} > 0$ favors upward motion ($m \to m+2$), lifting the weight above $x=0$.

In the continuum limit ($M$ large, $l$ small, $x = ml$), the discrete master equation approaches a Fokker--Planck / Smoluchowski equation for the probability density $\rho(x,t)$,
\begin{align}
\partial_t \rho(x,t) &= \frac{(2l)^2}{\tau}\partial_x\!\left(\partial_x - \big(\overbrace{\partial_x \SWM(x) + \betabath F}^{=\partial_x J}\big)\right)\rho(x,t) \nonumber \\
&= D\partial_x\!\left(\partial_x + \frac{1}{Ml^2}\big(x - \betabath F l^2 M\big)\right)\rho(x,t) \;, \label{Eq:FokkerPlanck}
\end{align}
with diffusion coefficient $D = (2l)^2/\tau$ (reflecting step size $2l$ in the continuum limit) and equilibrium center
\begin{equation}
\xbath = \betabath F l^2 M = -\betabath\Delta E L \;, \label{Eq:xbath}
\end{equation}
where $\SWM(x)$ is approximated quadratically around its maximum. The corresponding Green's function is Gaussian,
\begin{equation}
\rho(x_2|x_1;t) = \frac{1}{\sqrt{2\pi}\,\sigma(t)} \exp\!\left(-\frac{1}{2}\left(\frac{x_2 - \big(x_1 + \Delta x(t)\big)}{\sigma(t)}\right)^2\right) \;, \label{Eq:Greensdensity}
\end{equation}
with mean displacement and variance
\begin{align}
\Delta x(t) &= (\xbath - x_1)\left(1 - e^{-\alpha t}\right) \;, \label{Eq:Deltax} \\
\alpha &= \frac{D}{Ml^2} = \frac{4}{\tau M} \;, \\
\sigma^2(t) &= \left(1 - e^{-2\alpha t}\right) M l^2 \;. \label{Eq:alphasigma}
\end{align}
For even $M$ and even $m$, the discrete conditional probability follows by integrating over a cell of width $2l$ around $m_2 l$,
\begin{align}
p^{\text{Gauss}}(m_2|m_1,t) &= \int_{(m_2-1)l}^{(m_2+1)l} dx_2\,\rho(x_2|x_1;t) \nonumber \\
&\approx 2l\,\rho(m_2 l|m_1 l;t) \;, \label{Eq:GaussGreen3}
\end{align}
provided $\sigma(t)/l$ is sufficiently large (with boundary adjustments near $\pm M$).

As shown in \figref{Fig:7}, starting from $\xWM = x_1 = -20l$ ($\bWM > 0$) in a negative-temperature bath drives the mean position to $\xbath > 0$, shifting further upward as the bath becomes hotter ($\betabath$ more negative). In the long-time limit, the mean lift and extracted work are governed by the inverse-temperature difference $\Delta\beta = \bWM - \betabath$,
\begin{equation}
\lim_{t\to\infty} \Delta x(t) = \xbath - \xWM = L(\bWM - \betabath)\Delta E \;, \label{Eq:shift}
\end{equation}
and
\begin{align}
W &= \lim_{t\to\infty} -\Delta x(t) F \nonumber \\
&= (\underbrace{\bWM - \betabath}_{=\Delta\beta})\underbrace{\Delta E\,\Emax}_{\propto F^2} \;, \label{Eq:heatflow2}
\end{align}
scaling quadratically with the load force.

\subsection{The freely jointed chain as a thermodynamic donkey}
The freely jointed chain in a negative-temperature bath exhibits a counterintuitive response to an increased load (\figref{Fig:8}). Suppose the chain has equilibrated at $\xbath$ in the bath, so that $\bWM = \betabath$, and we take this as the new starting position,
\begin{equation}
\tilde{x}_{\text{WM}} = \xbath \;. \label{Eq:offsetnew}
\end{equation}
Increasing the magnitude of the downward force from $|F|$ to $|\tilde{F}| > |F|$ (for instance, by adding a mass or increasing $|\tilde{g}| > |g|$) increases the segment energy scale,
\begin{equation}
\Delta\tilde{E} = -l\tilde{F} > -lF = \Delta E \;. \label{Eq:DeltaEnew}
\end{equation}
Although the stronger force pulls downward in $-x$, the new equilibrium position shifts upward to
\begin{equation}
\tilde{x}_{\text{bath}} = -L\betabath\Delta\tilde{E} > -L\betabath\Delta E = \tilde{x}_{\text{WM}} \;,
\end{equation}
giving a positive net lift
\begin{equation}
\tilde{x}_{\text{bath}} - \tilde{x}_{\text{WM}} = L\betabath l(\tilde{F} - F) > 0 \;. \label{Eq:donkey}
\end{equation}

This ``donkey behavior'' follows directly from heat flow along a temperature gradient. Right after the force increase at $\tilde{x}_{\text{WM}} = \xbath$, the newly assigned inverse temperature of the WM is
\begin{align}
\tilde{\beta}_{\text{WM}} &= -\frac{1}{\Delta\tilde{E}}\left(\frac{\tilde{x}_{\text{WM}}}{L}\right) \nonumber \\
&= \betabath\frac{\Delta E}{\Delta\tilde{E}} = \betabath\frac{F}{\tilde{F}} > \betabath \;, \label{Eq:betacool}
\end{align}
which is cooler than the bath ($\betabath < 0$). Consequently, heat flows from the hotter bath into the cooler WM assembly, lifting the heavier load by
\begin{equation}
\tilde{x}_{\text{bath}} - \tilde{x}_{\text{WM}} = L(\tilde{\beta}_{\text{WM}} - \betabath)\Delta\tilde{E}
\end{equation}
and yielding additional work
\begin{align}
\tilde{W} &= -(\tilde{x}_{\text{bath}} - \tilde{x}_{\text{WM}})\tilde{F} \nonumber \\
&= (\tilde{\beta}_{\text{WM}} - \betabath)\Delta\tilde{E}\,\tilde{E}_{\text{max}} \;.
\end{align}
If an external work $W_{\text{ext}} = -\xbath(\tilde{F} - F) > 0$ is expended to place the added mass at $\xbath$, the net work gain is
\begin{equation}
\tilde{W} - W_{\text{ext}} = -\betabath(\tilde{F} - F)^2 lL \;,
\end{equation}
proportional to the square of the force increment.

\subsection{The freely jointed chain working as a cyclic machine in a single bath}
We can use this mechanism to construct a four-step cycle in a single negative-temperature bath (\figref{Fig:9}). In state~1, WMs carrying a baseline mass $\mu_0$ are at their bath equilibrium position, ${\xWM^{(1)}}_0 = {x_{\text{bath}}}_0$, with ${\bWM^{(1)}}_0 = \betabath$. In state~2, an additional load mass $\mu_{\text{load}}$ is attached at ${\xWM^{(2)}}_{\text{load}} = {\xWM^{(1)}}_0$. Because the total downward force becomes $F_{\text{load}} = -(\mu_0 + \mu_{\text{load}})g$, the WMs become cooler than the bath,
\begin{equation}
{\bWM^{(2)}}_{\text{load}} = \betabath \frac{\mu_0}{\mu_0 + \mu_{\text{load}}} > \betabath \;.
\end{equation}
Heat flows from the bath into the WM assembly ($Q_{\text{bath}} < 0$), lifting the combined mass to the new equilibrium position (state~3),
\begin{equation}
{x^{(2)}_{\text{bath}}}_{\text{load}} = -L\betabath l(\mu_0 + \mu_{\text{load}})g \;,
\end{equation}
and performing work
\begin{align}
W &= -Q_{\text{bath}} \nonumber \\
&= \left({x^{(2)}_{\text{bath}}}_{\text{load}} - {\xWM^{(1)}}_0\right)(\mu_0 + \mu_{\text{load}})g \nonumber \\
&= -L\betabath l\mu_{\text{load}}g(\mu_0 + \mu_{\text{load}})g \;.
\end{align}
In state~4, the load mass $\mu_{\text{load}}$ is removed, making the WMs hotter than the bath,
\begin{equation}
{\bWM^{(4)}}_0 = \betabath \frac{\mu_0 + \mu_{\text{load}}}{\mu_0} < \betabath \;.
\end{equation}
Heat now flows from the WMs back into the bath as the baseline mass returns to ${\xWM^{(1)}}_0$ (state~1), dissipating heat $Q_{\text{diss}} = -W_{\text{diss}} = \big({x^{(2)}_{\text{bath}}}_{\text{load}} - {\xWM^{(1)}}_0\big)\mu_0 g > 0$ with $|W_{\text{diss}}| < W$. Over a full cycle, the net work produced is
\begin{align}
W_{\text{cycle}} &= W + W_{\text{diss}} \nonumber \\
&= -L\betabath l(\mu_{\text{load}} g)^2 \;,
\end{align}
corresponding to an efficiency with respect to the absorbed heat $|Q_{\text{bath}}|$ of
\begin{equation}
W_{\text{cycle}} = \eta |Q_{\text{bath}}| \;, \qquad \eta = \frac{\mu_{\text{load}}}{\mu_0 + \mu_{\text{load}}} < 1 \;.
\end{equation}
Thus, a cyclic engine operating in a single negative-temperature bath cannot convert all absorbed heat into net cycle work, as part of the lifted energy must be re-dissipated into the bath when resetting the cycle.

\section{Summary and discussion}
How work can be extracted from a bath at negative temperature remains a subject of lively debate. Ramsey proposed extending the Kelvin--Planck formulation of the second law of thermodynamics \cite{19ramsey56thermodynamics}, whereas Struchtrup \cite{struchtrup2018work} argued that negative temperatures do not represent genuine equilibrium temperatures but rather non-equilibrium parameters. Here, we offer a resolution that neither modifies the second law nor denies negative temperatures. By assigning a canonical temperature to a working machine and comparing it with that of the bath, the temperature gradient dictates the direction of heat flow from hot to cold, and the heat transferred corresponds directly to the work performed. In our examples, a bounded energy spectrum enables the working machine to equilibrate with the bath and adopt its temperature.

Key ingredients of this approach are twofold: first, we consider an assembly of weakly coupled working machines rather than a single machine, defining a canonical temperature for the assembly by maximizing its Shannon entropy, which is also in the spirit of Jaynes's maximum-entropy approach to statistical mechanics \cite{jaynes1957information}. Second, we assume that a WM placed in a negative-temperature bath approaches a stationary state described by a canonical distribution $\propto e^{-\betabath E}$, as shown for systems at negative temperatures \cite{puglisi2017temperature,baldovin2021statistical}. Building on earlier work by Baldovin et al.\ \cite{baldovin2018langevin}, a general microscopic Langevin and Fokker--Planck description was derived via timescale separation for a slow heavy particle coupled to a fast bath at negative temperature. The existence of a steady state and detailed balance was demonstrated \cite{baldovin2019derivation}. Crucially, this stationary state is stable, meaning that an initial energy distribution relaxes toward it with decreasing Boltzmann $H$-function (increasing Shannon entropy), as shown recently in an elegant way in a kinetic model with a Stosszahlansatz by Lucente et al.\ \cite{lucente2025h}. Additionally it was shown in this model for the stationary state, that detailed balance. i.e. microreversibility, holds, i.e. the stationary state is also an equilibrium state. With these foundations, the dynamics of a WM in a negative-temperature bath is naturally treated within stochastic thermodynamics, obtaining transition rates from the principle of detailed balance and canonical fluctuations around the asymptotic equilibrium distribution.

Several aspects warrant further discussion. Usually, a working machine is viewed as a macroscopic device, making the assignment of a temperature seem counterintuitive. For microscopic machines, however, this picture is much more natural---especially when they operate collectively in assemblies (such as motors in muscle contraction), where assigning a temperature becomes physically transparent. One might also object that heat flow between two systems at different temperatures presupposes that both start in equilibrium with well-defined distributions. In our setup, a single WM starts in a definite state of energy $\Ewm$, and replicating it $N$ times with weak coupling produces an equilibrium assembly with mean energy $\Ewm$. Nevertheless, on long timescales the mean heat uptake---and thus the mean work performed---does not depend on the details of the initial distribution $p_0(E_{\text{i}})$, but only on its initial mean $\Ewm$. For an arbitrary initial distribution $p_0$ with mean $\Ewm$ in a bath at inverse temperature $\betabath$, the long-time joint probability factorizes,
\begin{align}
\lim_{t\to\infty} p(E_{\text{f}}, E_{\text{i}}; t) &= \left(\lim_{t\to\infty} p_{\betabath}(E_{\text{f}}|E_{\text{i}}; t)\right) p_0(E_{\text{i}}) \nonumber \\
&= p^{(\text{eq})}_{\betabath}(E_{\text{f}})\,p_0(E_{\text{i}}) \;,
\end{align}
so the mean work depends solely on the initial mean energy,
\begin{align}
W &= \langle E_{\text{f}} - E_{\text{i}} \rangle \nonumber \\
&= \langle E_{\text{f}} \rangle - \langle E_{\text{i}} \rangle \nonumber \\
&= \langle E \rangle^{(\text{eq})}_{\betabath} - \Ewm \;.
\end{align}

Finally, consider how the temperature of the WM assembly relates to that of the bath in determining heat flow. For the WM assembly, we obtain the canonical temperature by maximizing its Shannon energy entropy; for the bath, maximizing its Shannon entropy gives the same temperature as the standard Boltzmann entropy \cite{reif1965fundamentals}. Because thermal coupling between the WM assembly and the bath is weak, the total Shannon entropy of the combined system is given by the weighted sum of the individual Shannon entropies. Since canonical distributions satisfy $\partial_{\bar{E}} \Ssh = \beta$, heat flow from the hotter system (the bath) to the cooler system (the WM assembly) increases the combined Shannon entropy until both share the same canonical temperature at maximum entropy, i.e. heat flow is related to the canonical temperature difference. In this context it should be noted that the gradient of the canonical temperature as driving force of heat flow holds more generally than that of the Boltzmann temperature   \cite{bauer2022geometrically,bauer2024geometrically}.

\bibliography{literature}

\begin{appendix}

\section{Simulation}\label{Appendix1}
Stochastic transitions across the working machine's energy spectrum were simulated using the nearest-neighbor rates in Eq.~\eqref{Eq:transitionrates} and a small time step $dt$ (measured in units of $\tau$). For sufficiently small $dt/\tau$, the single-step probabilities for a hop up or down ($m' = m \pm 1$) are approximated by $(dt/\tau)\exp(\mp\betabath\Delta E/2)$, and the probability to remain in state $m$ is $1 - (dt/\tau)\,2\cosh(\betabath\Delta E/2)$, with reflecting boundary conditions at $m=1$ and $m=M$. Averaging over sample trajectories starting from the same initial state yields the time-dependent energy distribution.

\section{Analytical and numerical stochastic dynamics in the bath}\label{Appendix2}
Within the bath, the WM undergoes a stationary Markov process governed by the master equation for the state probabilities $p_{m'}(t)$,
\begin{equation}
\dot{p}_{m'} = \sum_{m=1}^{M} r_{m',m}\,p_m(t) \;,
\end{equation}
with $\sum_{m'} r_{m',m} = 0$ ensuring probability conservation. For nearest-neighbor transitions, Eq.~\eqref{Eq:transitionrates} gives upward rate $r_{m+1,m} = \tau^{-1}\exp(-\betabath\Delta E/2) \equiv c$ and downward rate $r_{m-1,m} = \tau^{-1}\exp(\betabath\Delta E/2) \equiv b$. This yields the tridiagonal $M \times M$ rate matrix
\begin{equation}
\mathbf{R} = \begin{pmatrix}
-c & b & 0 & \dots & 0 \\
c & -(c+b) & b & \dots & 0 \\
0 & c & -(c+b) & \ddots & 0 \\
\vdots & \vdots & \ddots & \ddots & b \\
0 & 0 & \dots & c & -b
\end{pmatrix} \;,
\end{equation}
describing discrete diffusion in a linear potential $E_m = (m-1)\Delta E$. Its $M$ eigenvalues $\lambda_k$, right eigenvectors $\mathbf{v}_k$, and left eigenvectors $\mathbf{w}_k$ are, for $k=0$,
\begin{align}
\lambda_0 &= 0 \;, \\
v_{0,m} &= \left(\frac{c}{b}\right)^{m-1} \;, \\
w_{0,m} &= 1 \;, \qquad m = 1, 2, \dots, M \;,
\end{align}
and for $k = 1, \dots, M-1$,
\begin{align}
\lambda_k &= -(b+c) + 2\sqrt{bc}\cos\!\left(\frac{k\pi}{M}\right) \;, \\
v_{k,m} &= \left(\sqrt{\frac{c}{b}}\right)^m \phi_k(m) \;, \\
w_{k,m} &= \left(\sqrt{\frac{b}{c}}\right)^{m-2} \phi_k(m) \;,
\end{align}
with
\begin{equation}
\phi_k(x) = \sin\!\left(\frac{kx\pi}{M}\right) - \sqrt{\frac{b}{c}}\sin\!\left(\frac{k(x-1)\pi}{M}\right) \;.
\end{equation}

Spectral decomposition of the propagator $\exp(\mathbf{R}t)$ gives the Green's function $p(m'|m;t)$ for a transition $m \to m'$ in time $t$,
\begin{equation}
\big(p(m'|m;t)\big) = \exp(\mathbf{R}t) = \sum_{k=0}^{M-1} e^{\lambda_k t}\,\frac{\mathbf{v}_k \mathbf{w}_k^T}{\mathbf{w}_k^T \mathbf{v}_k} \;,
\end{equation}
whose matrix elements take the form
\begin{align}
p(m'|m;t) &= p^{(\text{eq})}(m') + \frac{2}{M}\left(\sqrt{\frac{c}{b}}\right)^{m'-m} \nonumber \\
&\quad \cdot \sum_{k=1}^{M-1} e^{\lambda_k t}\,\frac{\phi_k(m')\phi_k(m)}{1 + \frac{b}{c} - 2\sqrt{\frac{b}{c}}\cos\!\left(\frac{k\pi}{M}\right)} \label{eq:meine_gleichung} \\
&= p^{(\text{eq})}(m') + f(m',m,t) \;, \nonumber
\end{align}
with the equilibrium distribution
\begin{equation}
p^{(\text{eq})}(m') = \left(\frac{c}{b}\right)^{m'-1} \frac{1 - \tfrac{c}{b}}{1 - \left(\tfrac{c}{b}\right)^M} \propto e^{-\betabath E_{m'}}
\end{equation}
and a multi-exponential relaxation term $f(m',m,t)$.

\section{Most probable distribution for WMs with internal degrees of freedom}\label{Appendix3}
Consider an assembly of $N$ working machines with energy occupation numbers $n(E)$. The number of ways to distribute the $N$ machines according to $\{n(E)\}$ is the multinomial coefficient $N!/\prod_E n(E)!$. With $\mathfrak{n}(E)$ microstates per energy level $E$, the total number of compatible microstates is
\begin{align}
\Omega[p] &= \frac{N!}{\prod_E n(E)!}\prod_E \mathfrak{n}(E)^{n(E)} \nonumber \\
&\approx \exp\!\big(N\,S_{\text{KL}}[p]\big) \;,
\end{align}
using Stirling's approximation, where
\begin{equation}
S_{\text{KL}}[p] = -\sum_E p(E)\ln\!\left(\frac{p(E)}{\mathfrak{n}(E)}\right)
\end{equation}
is the negative Kullback--Leibler divergence with $p(E) = n(E)/N$ and $\mathfrak{n}(E) = e^{\SWM(E)}$ (unnormalized, since any normalization factor only shifts $S_{\text{KL}}$ by a constant that drops out of the maximization). By equal a priori probabilities of a single microstate (principle of equal weight), the probability $\varpi[p]$ of observing distribution $p(E)$ satisfies $\varpi[p] \sim \Omega[p] \sim \exp\big(N\,S_{\text{KL}}[p]\big)$. Maximizing $S_{\text{KL}}[p]$ under the mean-energy constraint
\begin{equation}
\sum_E p(E) E = \bar{E} \label{Eq_app:constraint}
\end{equation}
with Lagrange multiplier $\beta$ via
\begin{equation}
\nabla_p \left(S_{\text{KL}}[p] - \beta \sum_E p(E) E\right)\bigg|_{p=p_c} = 0
\end{equation}
gives the canonical distribution
\begin{equation}
p_c(E) = \frac{1}{Z(\beta)}\,e^{\SWM(E) - \beta E} \;, \qquad Z(\beta) = \sum_E e^{\SWM(E) - \beta E} \;.
\end{equation}

\end{appendix}

\end{document}